\documentclass{emulateapj}
\usepackage{natbib}
\usepackage{amssymb}
\usepackage{amsmath}
\slugcomment{\emph{ApJ Letters}, in press, March 11, 2013}

\shorttitle{Chemistry of Impact-Generated Disks}

\shortauthors{Visscher and Fegley}

\begin{document}

\title{Chemistry of Impact-Generated Silicate Melt-Vapor Debris Disks}

\author{Channon Visscher}
\affil{Department of Space Studies, Southwest Research Institute, Boulder, CO, 80302, USA}
\author{Bruce Fegley, Jr.}
\affil{Planetary Chemistry Laboratory, Department of Earth \& Planetary Sciences and McDonnell Center for Space Sciences, Washington University in St.~Louis, St.~Louis, MO 63130, USA}

\begin{abstract}
In the giant impact theory for lunar origin, the Moon forms from material ejected by the impact into an Earth-orbiting disk.  Here we report the initial results from a silicate melt-vapor equilibrium chemistry model for such impact-generated planetary debris disks.  In order to simulate the chemical behavior of a two-phase (melt+vapor) disk, we calculate the temperature-dependent pressure and chemical composition of vapor in equilibrium with molten silicate from 2000 to 4000 K.  We consider the elements O, Na, K, Fe, Si, Mg, Ca, Al, Ti, and Zn for a range of bulk silicate compositions (Earth, Moon, Mars, eucrite parent body, angrites, and ureilites).  In general, the disk atmosphere is dominated by Na, Zn, and O$_{2}$ at lower temperatures ($< 3000$ K) and SiO, O$_{2}$, and O at higher temperatures.  The high-temperature chemistry is consistent for any silicate melt composition, and we thus expect abundant SiO, O$_{2}$, and O to be a common feature of hot, impact-generated debris disks.  In addition, the saturated silicate vapor is highly oxidizing, with oxygen fugacity ($f_{\textrm{O}_2}$) values (and hence H$_{2}$O/H$_{2}$ and CO$_{2}$/CO ratios) several orders of magnitude higher than those in a solar-composition gas.  High $f_{\textrm{O}_2}$ values in the disk atmosphere are found for any silicate composition because oxygen is the most abundant element in rock.  We thus expect high oxygen fugacity to be a ubiquitous feature of any silicate melt-vapor disk produced via collisions between rocky planets.
\end{abstract}


\section{Introduction}
In the giant-impact theory, the Moon formed as a result of an oblique collision between a Mars-sized impactor and the early Earth during its final stages of accretion \citep{hartmann1975,cameron1976}.  In the aftermath of this event, the Moon forms from the circumterrestrial disk of planetary debris  ejected by the impact.  The lunar composition thus provides constraints on the initial composition and subsequent chemical evolution of the protolunar disk.  Indeed, a longstanding strength of the giant impact scenario is its ability to account for a relatively Fe-poor, volatile-depleted Moon as an expected outcome of a large impact event with the differentiated proto-Earth \citep[e.g.,][]{hartmann1975,stevenson1987,canup2001,canup2008}.  

\begin{figure}
\begin{center}
\includegraphics[angle=0, width=0.45\textwidth]{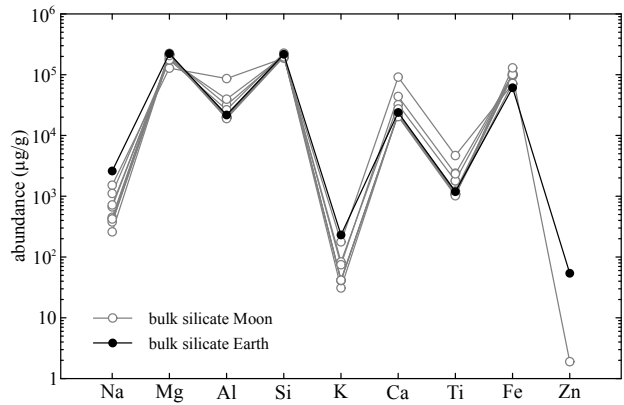} \end{center} \caption[Vapor Pressure]{\small{Elemental abundances in the bulk silicate Moon (gray lines with open symbols) and bulk silicate Earth (BSE; black lines with filled symbols) for elements included in the code calculations (in order of atomic number).  Lunar abundances are taken from multiple literature sources \citep[e.g.,][]{morgan1978,ringwood1979,taylor1982,wanke1982,ringwood1987,buck1980,delano1985,jones1989,oneill1991,warren2005,lodders2011}; BSE abundances are taken from \citet{palme2003}.}}
\label{fig: abundances} 
\end{figure}

Dynamical simulations of the giant impact have commonly found that the impactor provides most (60\%-80\%) of the disk material \citep[e.g.,][]{canup2001,canup2004,canup2008}. However, major element abundances in the bulk silicate Moon closely resemble those in Earth's mantle, albeit with slight enrichments in refractory elements and slight depletions in volatile elements \citep[e.g., see Figure \ref{fig: abundances};][]{ringwood1979,drake1986,ringwood1987}.  Moreover, the Earth and Moon possess apparently identical isotopic compositions of O, Cr, and Ti \citep{wiechert2001,lugmair1998,zhang2012}, which may require extensive re-mixing and equilibration between the early Earth and the impactor-dominated material of the disk in order to remove any Earth-Moon differences \citep{pahlevan2007}.  

Recent impact models suggest that a disk with a composition essentially equal to that of the post-impact planet can result for scenarios that include a rapidly rotating early Earth \citep{cuk2012} or a sufficiently large impactor \citep{canup2012}.  In these scenarios, Earth-Moon differences could presumably arise from the subsequent chemical evolution of the protolunar disk \citep[e.g.,][]{pahlevan2011,desch2011}.  For example, the observed fractionation of Zn isotopes in lunar samples suggests volatile loss during lunar formation \citep{paniello2012}.

In any case, the impact-generated debris is expected to form a two-phase silicate disk consisting of a melt layer surrounded by a vapor atmosphere \citep{stevenson1987,thompson1988,pahlevan2007,ward2012}.  However, the chemistry of the disk atmosphere, and its implications for the subsequent chemical evolution of the disk, remains poorly characterized, and only the evolution of the FeO/MgO ratio has been modeled \citep{pahlevan2011}.   Here we thus extend the models of \citet{fegley2012aas} and \citet{fegley2012} in order to explore the melt-vapor equilibrium chemistry of the protolunar disk.  In particular, we examine the chemical composition of the disk atmosphere, vapor pressure behavior of the silicate melt, and the prevailing oxygen fugacity of the disk.  

This letter is organized as follows. We describe our thermochemical equilibrium calculations for different melt compositions in \S2.  In \S3 we present our initial model results, including the chemical speciation of the saturated silicate vapor, the total vapor pressure over silicate melts, the oxygen fugacity, and H$_{2}$O/H$_{2}$ and CO$_{2}$/CO ratios in the disk.  We conclude in \S4 with a brief summary of our results and general implications for impact-generated silicate debris disks.  Although our models focus on conditions most relevant for lunar formation, our results are generally applicable for impact-generated disks following collisions between terrestrial planets.  The major gases that form in such collisions may thus provide chemical signatures of hot, impact-formed disks in extrasolar planetary systems.

\section{Computational Methods}
\subsection{Chemical Model}
We used the MAGMA code \citep{fegley1987,schaefer2004} to calculate the temperature-dependent chemical equilibrium composition of silicate melt and coexisting vapor.  The MAGMA code is a robust and rapid mass-balance, mass action algorithm based upon the ideal mixing of complex components (IMCC) model developed by Hastie et al.~\citep[see references in][]{fegley1987}.  A complete description of the MAGMA code, its method of calculation, and extensive validation against experimental data is given by \citet{fegley1987} and \citet{schaefer2004}.  Zinc thermochemistry was recently added to the code calculations in order to explore the chemical behavior and potential volatile loss of Zn-bearing gases from the protolunar disk atmosphere \citep[e.g.,][]{paniello2012}.  

The MAGMA code was used to simultaneously solve for activity coefficients in the melt, partial pressures in the gas phase, and melt-vapor equilibria for the elements Si, Mg, Fe, Al, Ca, Na, K, Ti, Zn, and O, considering 53 melt species and 32 gas species (see Appendix) from 1800-6000 K.  Neither H nor C are presently considered in the MAGMA code. However as noted by \citep{fegley2012}, the high temperature solubilities of H and C in molten silicates are expected to be low based on molecular dynamics simulations of CO$_{2}$ solubility and experimental measurements of water solubility versus temperature.  Lower temperatures were not considered because the MAGMA code is designed for completely molten systems above the liquidus temperature \citep{fegley1987}.  Here we focus on the results from 2000-4000 K, as this encompasses the expected range of temperatures in a two-phase protolunar disk from the photosphere to the midplane.  These temperatures are maintained over the $\sim$100 yr lifetime of the disk via a thermal feedback between viscous spreading and vaporization of the disk melt layer \citep{thompson1988,ward2012}.  Chemical equilibrium calculations are justified because equilibrium is reached faster than the disk cools or dissipates. \citet{fegley2012} calculated chemical lifetimes of $10^{-2}$ to $10^{-4}$ seconds for the oxidation of Si, Mg, or Fe atoms to their respective monoxide gases via reactions such as M + O$_{2}$ $\rightarrow$ MO + O in a silicate vapor at 2000 K, demonstrating that chemical equilibrium is quickly reached between the major Fe, Mg, O, and Si gases in the disk atmosphere.  

\begin{deluxetable}{lr@{.}lr@{.}lr@{.}lr@{.}lr@{.}lr@{.}l}[t*]
\tablefontsize{\footnotesize}
\tablewidth{0.48\textwidth}
\tablecaption{Bulk Silicate Compositions used in Model Calculations}
\tablehead{{oxide wt\%} & \multicolumn{2}{c}{BSE} & \multicolumn{2}{c}{Moon} & \multicolumn{2}{c}{Mars} & \multicolumn{2}{c}{EPB} & \multicolumn{2}{c}{Angrite} & \multicolumn{2}{c}{Ureilite}}
\startdata		
SiO$_{2}$ 		& 45&40 		& 44&60  	& 45&39 		& 46&26 		& 40&32 		& 43&65 		\\ 
MgO 				& 36&76 		& 35&10  	& 29&71 		& 31&55 		& 11&10 		& 39&98 		\\ 
Al$_{2}$O$_{3}$ 	&  4&48 		&   3&90 	&   2&89		& 3&27 		& 11&45 		&  0&37  	\\ 
TiO$_{2}$ 		&  0&21 		&   0&17 	&   0&14		& 0&16 		& 1&12 		& 0&10 		\\ 
FeO 				&  8&10 		&  12&40 	&  17&21		& 14&82 		& 20&10 		& 14&60 		\\ 
CaO 				&  3&65 		&   3&30 	&   2&36		& 2&58 		& 15&89 		& 1&23  		\\
Na$_{2}$O 		&  0&349		&  0&050 	&  0&983		& 0&110 		& 0&022 		& 0&065 		\\
K$_{2}$O 		&  0&031		&  0&004 	&  0&111		& 0&009 		& 0&007 		& 0&022		\\
ZnO				&  \multicolumn{2}{c}{6.7e-3} 	& \multicolumn{2}{c}{2.0e-4} & \multicolumn{2}{c}{7.6e-3}		& \multicolumn{2}{c}{4.0e-5}	& \multicolumn{2}{c}{5.0e-4}		& 0&027	\\
Total			&  99&0 	 	& 99&5  		&  98&8 		& 98&8 		& 100&0 		& 100&0 		\\
Mg mol\%			&  89&0 	 	& 83&3 	    &  75&5 		& 79&1		&  49&6 		&  83&0 \\
\enddata
\tablecomments{BSE: bulk silicate Earth \citep{palme2003}; Moon: bulk silicate Moon \citep{oneill1991}; Mars: bulk silicate Mars \citep{lodders2000,lodders2011}; EPB: Eucrite Parent Body, presumably 4 Vesta \citep[see][]{lodders2000}; Angrite is the unweighted mean composition of D'Orbigny, Sahara 99555, Angra dos Reis, LEW 86010, and LEW 87051 angrites computed from \citet{warren1990,warren1995,mittlefehldt2002}; Ureilite is the unweighted median composition of the ALH 84136, ALHA 77257, Dinngo Pup Donga, Dyalpur, Havero, Goalpara, North Haig, Novo Urei, and PCA 82506 ureilites computed from \citet{ringwood1960,vdovykin1970,wiik1972,jarosewich1990,klock1998}.  All abundances are renormalized to 100\% for code calculations.}
\label{tab: compositions}
\end{deluxetable}

\begin{figure*}[ht]
\begin{center}
\includegraphics[angle=0, width=0.95\textwidth]{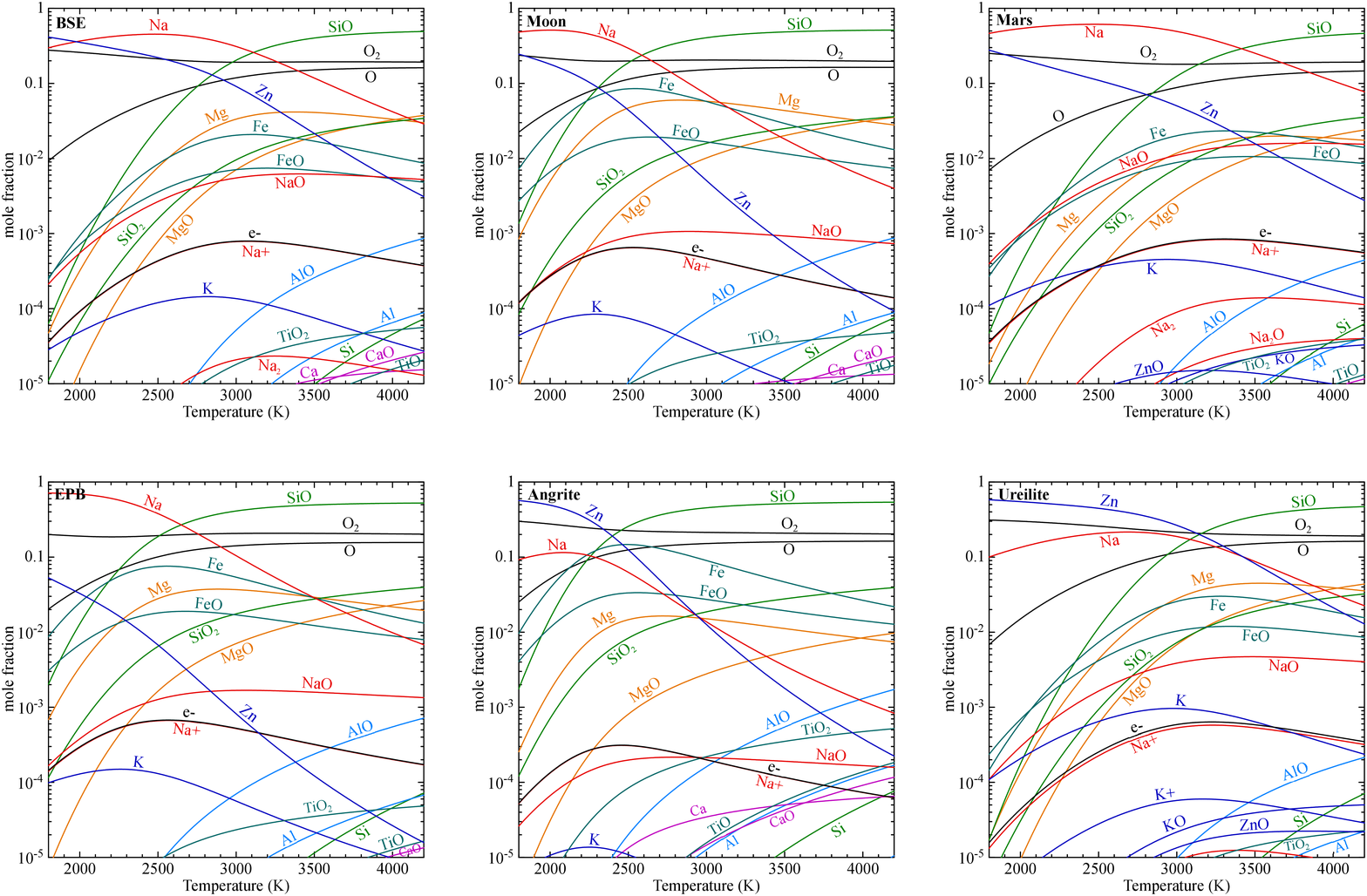} \end{center} \caption[Magma Vapor Chemistry]{\small{Chemical composition of the saturated silicate vapor as a function of temperature, in equilibrium with melt compositions corresponding to a) bulk silicate Earth (BSE), b) bulk silicate Moon, c) bulk silicate Mars, d) the Eucrite parent body (EPB), e) angrites, and f) ureilites.  Table \ref{tab: compositions} lists the oxide abundances for each case.}}
\label{fig: speciation} 
\end{figure*}

\subsection{Melt Composition}
We considered several different melt compositions, including bulk silicate compositions for the Earth, Moon, and Mars.  Dating by three relative chronometers ($^{182}$Hf-$^{182}$W, $^{26}$Al-$^{26}$Mg, and $^{53}$Mn-$^{53}$Cr) indicates that the parent body of the eucrite, howardite, and diogenite achondrites differentiated during the first 1-10 Ma of solar system history \citep{jacobsen2008,trinquier2008,schiller2011}.  We thus also included the major groups of differentiated achondrite meteorites in our modeling by considering the average composition of the eucrite parent body (EPB; presumably 4 Vesta), angrites, and ureilites.  The oxide abundances and data sources for each melt composition are listed in Table \ref{tab: compositions}.  

To a rough approximation, the bulk silicate Earth and bulk silicate Moon can be assumed to represent the initial and final composition, respectively, of the protolunar disk material.  We studied a range of bulk lunar compositions from the literature (e.g., see Figure \ref{fig: abundances}), and adopted \citet{oneill1991} values as our nominal bulk silicate Moon composition for the model results reported here.  

As we discuss below, differences in the model results are primarily caused by differing abundances of volatile elements.  For example, the bulk silicate Earth has Na, K, and Zn abundances of 2590, 260, and 54 $\mu$g/g, respectively \citep{palme2003}, whereas the corresponding lunar abundances are generally a magnitude lower, at 260, 31, and 1.9 $\mu$g/g, respectively \citep{oneill1991}.  Model lunar compositions with intermediate abundances of volatiles yield chemical trends and vapor pressures intermediate to the ``end-member'' results for bulk silicate Moon \citep{oneill1991} and bulk silicate Earth \citep{palme2003}.  Moreover, model compositions with relatively lower Fe abundances and higher Mg numbers \citep[e.g.,][]{warren2005} give similar results as our nominal lunar model because Fe and Mg have less influence on the overall chemical behavior of the disk atmosphere.

\begin{figure*}
\begin{center}
\includegraphics[angle=0, width=0.75\textwidth]{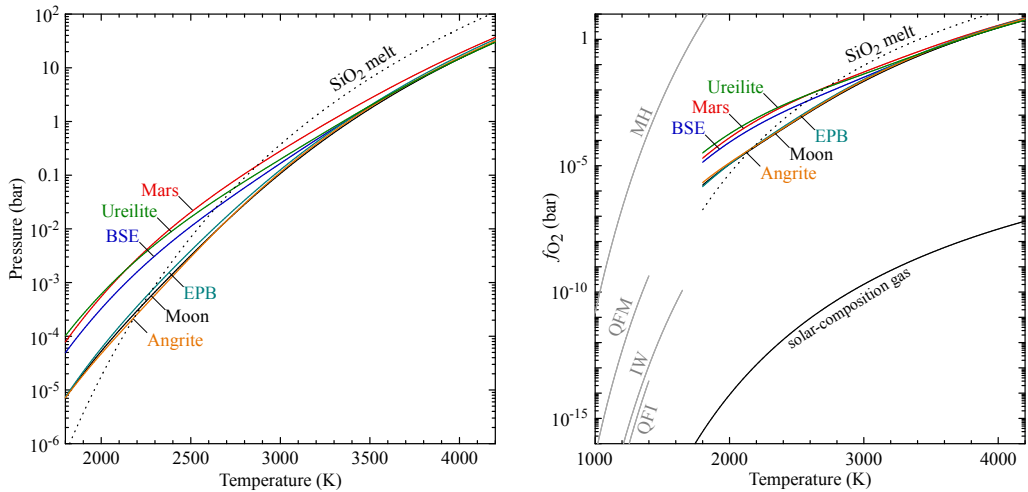} \end{center} \caption[Vapor Pressure]{\small{Model results showing a) total vapor pressure of the saturated silicate vapor for different melt compositions, and b) oxygen fugacity as a function of temperature.  Also shown are results for pure molten SiO$_{2}$ \citep[e.g.,][]{thompson1988,ward2012}, and $f_{\textrm{O}_{2}}$ for a solar-composition gas \citep{lodders2003} and common oxygen fugacity buffers. MH: magnetite-hematite, QFM: quartz-fayalite-magnetite, IW: iron-w\"{u}stite, QFI: quartz-fayalite-iron.}}
\label{fig: pressure} 
\end{figure*}

\section{Results}

\subsection{Chemical speciation of saturated silicate vapor}
The chemistry of the saturated silicate vapor (i.e., the disk atmosphere) is shown in Figure \ref{fig: speciation} for different melt compositions over the temperature range 2000-4000 K.  A number of general trends are observed for all silicate melt compositions.  At high temperatures ($>3000$ K), the disk atmosphere is dominated by SiO, O$_{2}$, and O, whereas more volatile species such as Na and Zn tend to dominate the silicate vapor at lower temperatures.  The mole fraction abundance of O$_{2}$ is relatively temperature-independent, and O$_{2}$ comprises 20-30\% of the silicate vapor for all melt compositions.

The chemical behavior of a given element in the silicate vapor is a function of both its abundance in the melt and its saturation vapor pressure behavior over the melt.  By definition, volatile elements are characterized by higher vapor pressures.  Thus, at lower temperatures, refractory elements preferentially remain in the melt whereas volatile elements more readily enter the vapor, so that Na- and Zn-bearing species dominate the disk atmosphere.  As temperatures increase, however, more disk material is vaporized and even refractory elements (e.g., Ca, Ti, Al) enter the vapor in increasing amounts.  

At high temperatures, the relative elemental abundances in the system play a larger role in determining the vapor composition, so that silicon (the most abundant element in the system after oxygen) will dominate the vapor as SiO, followed by O$_{2}$, O, and the moderately volatile Mg- and Fe-bearing gases, which typically comprise 1\%-10\% of the vapor \citep[e.g., Fe-oxides in the melt dissociate into Fe, FeO, O, and O$_{2}$ in the vapor;][]{kazenas1995}.  The temperature of the transition between a SiO-dominated regime and a Na- and/or Zn-dominated regime depends upon the relative abundances of the volatiles in the melt.  For example, for bulk compositions with relatively higher abundances of volatiles (BSE, Mars, Ureilite) this transition occurs between 3000-3500 K.  For melt compositions with lower volatile abundances (Moon, EPB, Angrite), this transition occurs at temperatures near 2500 K.  Moreover, as we discuss below, these basic differences in volatile content and disk chemistry also strongly influence the total vapor pressure and oxygen fugacity of the disk atmosphere.

\subsection{Total vapor pressure of saturated silicate vapor}
The vapor pressure of the saturated silicate vapor in equilibrium with the melt is an important parameter that describes the thermal structure of the disk.  \citet{thompson1988} and \citet{ward2012} used the Clausius-Clapeyron relation to describe the pressure-temperature profile of a two-phase disk,
\begin{equation}\label{eq: clausius}
P = P_o e^{-(T_o/T)},
\end{equation}
and adopt \citet{krieger1965} data for vapor pressure over pure molten SiO$_{2}$, using values of $P_o = 2.23 \times 10^{8}$ bar and $T_o = 6.03\times10^4$ K in the above expression.  \citet{fegley2012} re-examined the vapor pressure behavior of molten silica up to 6000 K by considering chemical equilibria between molten silica and all species in the saturated vapor using the MAGMA code, giving the expression:
\begin{equation}
\log P_{vap}(\textrm{SiO}_{2}, liq) = 8.203 - 25898.9/T,
\end{equation}
for $P$ in bars and $T$ in K \citep[cf.][]{melosh2007}.  However, as noted by \citet{fegley2012}, the lunar disk is not pure silica but instead a mixture of multiple oxide components.  We therefore calculated the total saturation vapor pressure in equilibrium with multicomponent silicate melts.  The results are shown in Figure \ref{fig: pressure}a.

\begin{figure*}[t]
\begin{center}
\includegraphics[angle=0, width=0.75\textwidth]{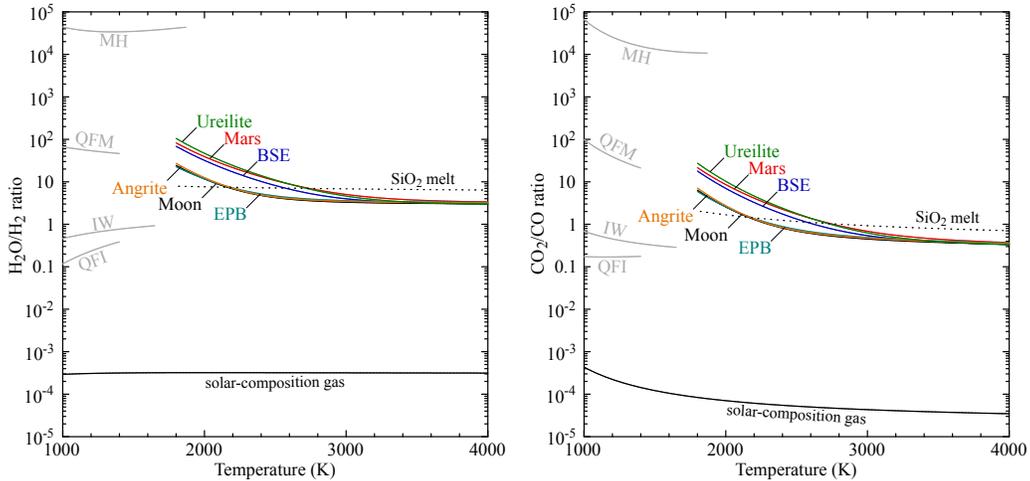} \end{center} \caption[Fugacity]{\small{Model results showing a) H$_{2}$O/H$_{2}$ ratios, and b) CO$_{2}$/CO ratios in saturated silicate vapor as a function of temperature.  The corresponding ratios for pure molten SiO$_{2}$, a solar-composition gas \citep{lodders2003}, and common oxygen fugacity buffers are shown for comparison.}}
\label{fig: fugacity} 
\end{figure*}

The saturation vapor pressure over a multicomponent melt is the sum of the partial pressures of all gases in equilibrium with a melt containing multiple oxides,
\begin{equation}
P_T = \Sigma P_i,
\end{equation}
where $i$ is a gas species (see Appendix) with a partial pressure $P_i$ that is a function of the melt-vapor equilibria.  For example, the net thermochemical reaction for SiO vapor can be written as
\begin{equation}\label{rxn: SiO}
\textrm{SiO}_{2}(\textrm{liq}) = \textrm{SiO(g)} + 0.5\textrm{O}_{2}\textrm{(g)}.
\end{equation}
The partial pressure of SiO in the vapor is given by
\begin{equation}
P_{\textrm{SiO}} = K_{\ref{rxn: SiO}}a_{\textrm{SiO}_{2}}f_{\textrm{O}_{2}}^{-0.5}
\end{equation}
where $K_{\ref{rxn: SiO}}$ is the temperature-dependent equilibrium constant, $f_{\textrm{O}_{2}}$ is the oxygen fugacity, and $a_{\textrm{SiO}_{2}}$ is the thermodynamic activity of liquid SiO$_{2}$ in a multicomponent melt. For volatile-bearing silicate melts, the saturated vapor is dominated by Na- and Zn-bearing gases at lower temperatures and by SiO at higher temperatures.  Melt compositions with relatively higher volatile element abundances (e.g., BSE, Mars, Ureilite) therefore tend to have relatively higher saturation vapor pressures ($P_{T}$) at lower temperatures, as is demonstrated in the vapor pressure curves in Figure \ref{fig: pressure}a.  The vapor pressure (bar) as a function of temperature is approximated by
\begin{equation}
\log P_{vap}(\textrm{BSE}) \approx -51.17 - 1048.8/T + 14.60 \log T
\end{equation}
for a BSE-composition melt \citep{palme2003}, and
\begin{equation}
\log P_{vap}(\textrm{BSM}) \approx -41.45 - 4052.3/T + 14.90 \log T
\end{equation}
for a BSM-composition melt \citep{oneill1991}, over the range of temperatures relevant for the disk atmosphere (2000-4000 K).

Our models do not currently include H$_{2}$O, and a full treatment of water solubility in high-temperature ($>1800$ K) silicate melts is beyond the scope of this letter.  However, we are able to make a number of general predictions regarding the behavior of water in the disk.  Because of the relatively low solubility of water in magma (especially for high-temperature, low-pressure post-impact conditions), most H$_{2}$O in the disk will remain in the gas phase.  Thus, for even minor amounts of water in the disk, we expect H$_{2}$O to be a major gas in the disk atmosphere (for similar reasons that Na and Zn dominate at lower temperatures), leading to an increase in the total vapor pressure of the disk. \citet{fegley2012} reached a similar conclusion for the post-impact silicate vapor atmosphere of the Earth.  We therefore consider any vapor pressure estimates for anhydrous melts (whether pure molten SiO$_{2}$ or silicate magmas) to represent lower limits for the thermal structure of disk atmospheres.

\subsection{Oxygen fugacity of saturated silicate vapor}
The equilibrium chemistry between the silicate magma and coexisting vapor determines the overall oxygen fugacity ($f_{\textrm{O}_{2}}$) of the system, given by
\begin{equation}
f_{\textrm{O}_{2}} = \gamma_{\textrm{O}_{2}}P_{\textrm{O}_{2}}=\gamma_{\textrm{O}_{2}}X_{\textrm{O}_{2}}P_T,
\end{equation}
where $X_{\textrm{O}_{2}}$ is the mole fraction abundance (see Figure \ref{fig: speciation}), $P_{\textrm{O}_{2}}$ is the partial pressure, and $\gamma_{\textrm{O}_{2}}$ is the fugacity coefficient of O$_{2}$ in a gas of total pressure $P_T$ (see Figure \ref{fig: pressure}a).   Over the range of temperatures and pressures considered here, $\gamma_{\textrm{O}_{2}}\approx 1$, so we take $f_{\textrm{O}_{2}} \approx P_{\textrm{O}_{2}} \approx X_{\textrm{O}_{2}}P_T$ from our model results.  Figure \ref{fig: pressure}b shows the oxygen fugacity as a function of temperature for different silicate melt compositions, including pure molten silica.   Also shown for comparison are $f_{\textrm{O}_{2}}$ values for some common oxygen fugacity mineral buffers and for a solar-composition gas \citep{lodders2003}.  Differences between model compositions at lower temperatures are again caused by the relative abundances of volatiles (e.g., Na, Zn) in each melt, and the $f_{\textrm{O}_{2}}$ values converge at high temperatures where SiO is the dominant species in the disk atmosphere.  For each case the saturated silicate vapor is highly oxidizing, with $f_{\textrm{O}_{2}}$ values several orders of magnitude higher than those expected in a solar-composition gas.  In fact, we expect very high $f_{\textrm{O}_{2}}$ values for \emph{any} silicate melt composition because oxygen is the most abundant element in rock.  

Although H$_{2}$O and H$_{2}$ are not included in the magma models, the prevailing oxygen fugacity --- which is controlled by silicate melt-vapor equilibrium chemistry --- can be used to constrain H$_{2}$O/H$_{2}$O ratios in the disk.  For the net thermochemical reaction,
\begin{equation}\label{rxn: H2O}
\textrm{H}_2 + 0.5\textrm{O}_{2} = \textrm{H}_{2}\textrm{O},
\end{equation}
the equilibrium constant expression can be rearranged to give the H$_{2}$O/H$_{2}$ ratio as a function of the oxygen fugacity ($f_{\textrm{O}_{2}}$) and the temperature-dependent equilibrium constant $K_{\ref{rxn: H2O}}$:
\begin{equation}
\textrm{H}_{2}\textrm{O}/\textrm{H}_{2} = K_{\ref{rxn: H2O}}f_{\textrm{O}_{2}}^{0.5},
\end{equation}
where $f_{\textrm{O}_{2}}$ is given by the silicate melt-vapor equilibrium chemistry.

In the same way, the prevailing oxygen fugacity of the saturated silicate vapor can also be used to determine the relative abundances of CO$_{2}$ and CO via the reaction,
\begin{equation}\label{rxn: CO2}
\textrm{CO} + 0.5\textrm{O}_{2} = \textrm{CO}_{2},
\end{equation}
where the CO$_{2}$/CO ratio is given by
\begin{equation}
\textrm{CO}_{2}/\textrm{CO} = K_{\ref{rxn: CO2}}f_{\textrm{O}_{2}}^{0.5}.
\end{equation}
The H$_{2}$O/H$_{2}$ and CO$_{2}$/CO ratios as a function of temperature are shown in Figures \ref{fig: fugacity}a and \ref{fig: fugacity}b, respectively.  Because these ratios are determined by the oxygen fugacity in the system, they are again significantly higher for an impact-generated disk atmosphere than for a solar-composition gas.  For the silicate melts, the H$_{2}$O/H$_{2}$ ratio ranges from $\sim20$ (Moon, Angrite, EPB) to $\sim100$ (Ureilite, Earth, BSE) at 1800 K to a nearly constant ratio of $\sim3$ at higher temperatures ($> 3000$ K) for all silicate compositions.  Similar behavior is observed for CO$_{2}$/CO ratios, with values from $\sim6$ (Moon, Angrite, EPB) to $\sim30$ (Ureilite, Mars, BSE) at 1800 K to a nearly constant ratio of $\sim0.3$ at higher temperatures.  For comparison,  H$_{2}$O/H$_{2}\sim0.0003$ and CO$_{2}$/CO $\lesssim0.0001$ in a solar composition gas \citep[][]{lodders2003}.

\section{Summary and Conclusions}

We used thermochemical equilibrium calculations to model the chemistry of the circumterrestrial melt-vapor disk resulting from the Moon-forming impact.  Our results show that the silicate vapor in the disk is dominantly composed of SiO, O$_{2}$, O at higher temperatures and Na, Zn, and O$_{2}$ at lower temperatures.  Although similar chemical trends are observed for all model compositions, the relative abundances of volatile elements in the melt (particularly Na and Zn) strongly influence the chemical composition, vapor pressure, and oxygen fugacity of the disk atmosphere at lower temperatures.  Although not yet incorporated into our models, we expect similar chemical behavior from other volatile elements (e.g., Rb, Cs, Pb, Sn, etc.) which will readily enter the disk atmosphere.  The chemical evolution of the disk due to volatile loss \citep[e.g., for Zn as reported by][]{paniello2012} will be explored elsewhere.

For any silicate composition, the silicate vapor in equilibrium with the melt is significantly more oxidizing (by several orders of magnitude) than a solar-composition gas.  High $f_{\textrm{O}_{2}}$ values in turn indicate high CO$_{2}$/CO ratios and high H$_{2}$O/H$_{2}$ ratios, suggesting that only minor amounts of H$_{2}$ and H will be produced by H$_{2}$O dissociation over the lifetime of the disk.  Because oxygen is the most abundant element in rock, high oxygen fugacity is a key feature of the protolunar disk, or any impact-generated debris disk that forms via collisions between rocky planets.  High abundances of SiO, O$_{2}$, and O may thus serve as chemical fingerprints of hot, impact-formed disks in extrasolar planetary systems.\\  

We thank F. Moynier and R. Canup for helpful discussions.  Work by CV was supported by the Center for Lunar Origin and Evolution (CLOE) as part of the NASA Lunar Science Institute (NLSI).  Work by BF was supported by NASA cooperative agreement NNX09AG69A with the NASA Ames Research Center and by grants from the NSF Astronomy Program.

\appendix
\section{Species included in MAGMA calculations} 
\small
\sloppy
\noindent \emph{Melt Species}: 
SiO$_{2}$,
MgO,
FeO,
Fe$_{2}$O$_{3}$,
Fe$_{3}$O$_{4}$,
CaO,
Al$_{2}$O$_{3}$,
TiO$_{2}$,
Na$_{2}$O,
K$_{2}$O,
ZnO,
MgSiO$_{3}$,
Mg$_{2}$SiO$_{4}$,
MgAl$_{2}$O$_{4}$,
MgTiO$_{3}$,
MgTi$_{2}$O$_{5}$,
Mg$_{2}$TiO$_{4}$,
Al$_{6}$Si$_{2}$O$_{1}$$_{3}$,
CaAl$_{2}$O$_{4}$,
CaAl$_{4}$O$_{7}$,
Ca$_{1}$$_{2}$Al$_{1}$$_{4}$O$_{3}$$_{3}$,
CaSiO$_{3}$,
CaAl$_{2}$Si$_{2}$O$_{8}$,
CaMgSi$_{2}$O$_{6}$,
Ca$_{2}$MgSi$_{2}$O$_{7}$,
Ca$_{2}$Al$_{2}$SiO$_{7}$,
CaTiO$_{3}$,
Ca$_{2}$SiO$_{4}$,
CaTiSiO$_{5}$,
FeTiO$_{3}$,
Fe$_{2}$SiO$_{4}$,
FeAl$_{2}$O$_{4}$,
CaAl$_{1}$$_{2}$O$_{1}$$_{9}$,
Mg$_{2}$Al$_{4}$Si$_{5}$O$_{1}$$_{8}$,
Na$_{2}$SiO$_{3}$,
Na$_{2}$Si$_{2}$O$_{5}$,
NaAlSiO$_{4}$,
NaAlSi$_{3}$O$_{8}$,
NaAlO$_{2}$,
Na$_{2}$TiO$_{3}$,
NaAlSi$_{2}$O$_{6}$,
K$_{2}$SiO$_{3}$,
K$_{2}$Si$_{2}$O$_{5}$,
KAlSiO$_{4}$,
KAlSi$_{3}$O$_{8}$,
KAlO$_{2}$,
KAlSi$_{2}$O$_{6}$,
K$_{2}$Si$_{4}$O$_{9}$,
KCaAlSi$_{2}$O$_{7}$,
Zn$_{2}$SiO$_{4}$,
ZnTiO$_{3}$,
Zn$_{2}$TiO$_{4}$,
ZnAl$_{2}$O$_{4}$\\

\noindent \emph{Gas Species}:
O,
O$_{2}$,
Mg,
MgO,
Si,
SiO,
SiO$_{2}$,
Fe,
FeO,
Al,
AlO,
AlO$_{2}$,
Al$_{2}$O,
Al$_{2}$O$_{2}$,
Ca,
CaO,
Na,
Na$_{2}$,
NaO,
Na$_{2}$O,
Na$^{+}$,
K,
K$_{2}$,
KO,
K$_{2}$O,
K$^{+}$,
Ti,
TiO,
TiO$_{2}$,
Zn,
ZnO,
e$^{-}$


\normalsize

\begin{thebibliography}{50}
\expandafter\ifx\csname natexlab\endcsname\relax\def\natexlab#1{#1}\fi

\bibitem[{{Buck} \& {Toks\"{o}z}(1980)}]{buck1980}
{Buck}, W.~R., \& {Toks\"{o}z}, M.~N. 1980, in Lunar and Planetary Science
  Conference Proceedings, 11th, 2043--2058

\bibitem[{{Cameron} \& {Ward}(1976)}]{cameron1976}
{Cameron}, A.~G.~W., \& {Ward}, W.~R. 1976, LPSC 7,120

\bibitem[{{Canup}(2004)}]{canup2004}
{Canup}, R.~M. 2004, Icarus, 168, 433

\bibitem[{{Canup}(2008)}]{canup2008}
---. 2008, Icarus, 196, 518

\bibitem[{{Canup}(2012)}]{canup2012}
---. 2012, Science, 338, 1052

\bibitem[{{Canup} \& {Asphaug}(2001)}]{canup2001}
{Canup}, R.~M., \& {Asphaug}, E. 2001, \nat, 412, 708

\bibitem[{{{\'C}uk} \& {Stewart}(2012)}]{cuk2012}
{{\'C}uk}, M., \& {Stewart}, S.~T. 2012, Science, 338, 1047

\bibitem[{{Delano}(1985)}]{delano1985}
{Delano}, J.~W. 1985, LPSC XVI, 179--180

\bibitem[{{Desch} \& {Taylor}(2011)}]{desch2011}
{Desch}, S.~J. \& {Taylor}, G.~J., 2011, {LPSC XLII}, 2005

\bibitem[{{Drake}(1986)}]{drake1986}
{Drake}, M.~J. 1986, in Origin of the Moon, ed. W.~K. {Hartmann}, R.~J.
  {Phillips}, \& G.~J. {Taylor}, 105--124

\bibitem[{{Fegley} \& {Cameron}(1987)}]{fegley1987}
{Fegley}, B., \& {Cameron}, A.~G.~W. 1987, Earth and Planetary Science Letters,
  82, 207

\bibitem[{{Fegley} {et~al.}(2012){Fegley}, {Schaefer}, \&
  {Lodders}}]{fegley2012aas}
{Fegley}, B., {Schaefer}, L., \& {Lodders}, K. 2012, AAS Meeting Abstracts 219, 301.01

\bibitem[{{Fegley} \& {Schaefer}(2012)}]{fegley2012}
{Fegley}, Jr, B., \& {Schaefer}, L. 2012, ArXiv e-prints

\bibitem[{{Hartmann} \& {Davis}(1975)}]{hartmann1975}
{Hartmann}, W.~K., \& {Davis}, D.~R. 1975, Icarus, 24, 504

\bibitem[{Jacobsen {et~al.}(2008)Jacobsen, Ranen, Petaev, Remo, O'Connell, \&
  Sasselov}]{jacobsen2008}
Jacobsen, S.~B., Ranen, M.~C., Petaev, M.~I., Remo, J.~L., O'Connell, R.~J., \&
  Sasselov, D.~D. 2008, Phil. Trans. Roy. Soc. A, 366, 4129

\bibitem[{{Jarosewich}(1990)}]{jarosewich1990}
{Jarosewich}, E. 1990, Meteoritics, 25, 323

\bibitem[{{Jones} \& {Delano}(1989)}]{jones1989}
{Jones}, J.~H., \& {Delano}, J.~W. 1989, \gca, 53, 513

\bibitem[{{Kazenas} \& {Tagirov}(1995)}]{kazenas1995}
{Kazenas}, E.~K. \& {Tagirov}, V.~K. 1995, Metally, 2, 31

\bibitem[{{Kl\"{o}ck} {et~al.}(1998){Kl\"{o}ck}, {Nakamura}, {Maetz}, \&
  {Arndt}}]{klock1998}
{Kl\"{o}ck}, W., {Nakamura}, K., {Maetz}, M., \& {Arndt}, P. 1998, Meteoritics
  and Planetary Science Supplement, 33, 84

\bibitem[{{Krieger}(1965)}]{krieger1965}
{Krieger}, F.~J. 1965, Rand Corporation Memorandum RM-4804-PR

\bibitem[{{Lodders}(2000)}]{lodders2000}
{Lodders}, K. 2000, \ssr, 92, 341

\bibitem[{{Lodders}(2003)}]{lodders2003}
---. 2003, Astrophysical Journal, 591, 1220

\bibitem[{{Lodders} \& {Fegley}(2011)}]{lodders2011}
{Lodders}, K., \& {Fegley}, B. 2011, Chemistry of the Solar System (Cambridge:
  Royal Society of Chemistry)

\bibitem[{{Lugmair} \& {Shukolyukov}(1998)}]{lugmair1998}
{Lugmair}, G.~W., \& {Shukolyukov}, A. 1998, \gca, 62, 2863

\bibitem[{{Melosh}(2007)}]{melosh2007}
{Melosh}, H.~J. 2007, Meteoritics and Planetary Science, 42, 2079

\bibitem[{{Mittlefehldt} {et~al.}(2002){Mittlefehldt}, {Killgore}, \&
  {Lee}}]{mittlefehldt2002}
{Mittlefehldt}, D.~W., {Killgore}, M., \& {Lee}, M.~T. 2002, Meteoritics and
  Planetary Science, 37, 345

\bibitem[{{Morgan} {et~al.}(1978){Morgan}, {Hertogen}, \&
  {Anders}}]{morgan1978}
{Morgan}, J.~W., {Hertogen}, J., \& {Anders}, A. 1978, Moon and Planets, 18,
  465

\bibitem[{{O'Neill}(1991)}]{oneill1991}
{O'Neill}, H.~S.~C. 1991, \gca, 55, 1135

\bibitem[{{Pahlevan} \& {Stevenson}(2007)}]{pahlevan2007}
{Pahlevan}, K., \& {Stevenson}, D.~J. 2007, Earth and Planetary Science
  Letters, 262, 438
  
\bibitem[{{Pahlevan} {et~al.}(2011){Pahlevan}, {Stevenson}, \&
  {Eiler}}]{pahlevan2011}
{Pahlevan}, K., {Stevenson}, D.~J., \& {Eiler}, J.~M., 2011. 
{Earth and Planetary Science Letters}, 301, 433--443.

\bibitem[{{Palme} \& {O'Neill}(2003)}]{palme2003}
{Palme}, H., \& {O'Neill}, H.~S.~C. 2003, Treatise on Geochemistry, 2, 1

\bibitem[{{Paniello} {et~al.}(2012){Paniello}, {Day}, \&
  {Moynier}}]{paniello2012}
{Paniello}, R.~C., {Day}, J.~M.~D., \& {Moynier}, F. 2012, \nat, 490, 376

\bibitem[{{Ringwood}(1960)}]{ringwood1960}
{Ringwood}, A.~E. 1960, \gca, 20, 1

\bibitem[{{Ringwood}(1979)}]{ringwood1979}
---. 1979, {Origin of the earth and moon} (New York: Springer-Verlag)

\bibitem[{{Ringwood} {et~al.}(1987){Ringwood}, {Seifert}, \&
  {Waenke}}]{ringwood1987}
{Ringwood}, A.~E., {Seifert}, S., \& {Waenke}, H. 1987, Earth and Planetary
  Science Letters, 81, 105

\bibitem[{{Schaefer} \& {Fegley}(2004)}]{schaefer2004}
{Schaefer}, L., \& {Fegley}, B. 2004, Icarus, 169, 216

\bibitem[{{Schiller} {et~al.}(2011){Schiller}, {Baker}, {Creech}, {Paton},
  {Millet}, {Irving}, \& {Bizzarro}}]{schiller2011}
{Schiller}, M., {Baker}, J., {Creech}, J., {Paton}, C., {Millet}, M.-A.,
  {Irving}, A., \& {Bizzarro}, M. 2011, \apjl, 740, L22

\bibitem[{{Stevenson}(1987)}]{stevenson1987}
{Stevenson}, D.~J. 1987, Annual Review of Earth and Planetary Sciences, 15, 271

\bibitem[{{Taylor}(1982)}]{taylor1982}
{Taylor}, S.~R. 1982, Physics of the Earth and Planetary Interiors, 29, 233

\bibitem[{{Thompson} \& {Stevenson}(1988)}]{thompson1988}
{Thompson}, C., \& {Stevenson}, D.~J. 1988, \apj, 333, 452

\bibitem[{{Trinquier} {et~al.}(2008){Trinquier}, {Birck}, {All{\`e}gre},
  {G{\"o}pel}, \& {Ulfbeck}}]{trinquier2008}
{Trinquier}, A., {Birck}, J.-L., {All{\`e}gre}, C.~J., {G{\"o}pel}, C., \&
  {Ulfbeck}, D. 2008, \gca, 72, 5146

\bibitem[{{Vdovykin}(1970)}]{vdovykin1970}
{Vdovykin}, G.~P. 1970, \ssr, 10, 483

\bibitem[{{W\"{a}nke} \& Dreibus(1982)}]{wanke1982}
{W\"{a}nke}, H., \& Dreibus. 1982, in Tidal Friction and the Earth's Rotation
  II, ed. P.~{Brosche} \& J.~{S\"{u}ndermann} (Berlin: Springer-Verlag),
  322--344

\bibitem[{{Ward}(2012)}]{ward2012}
{Ward}, W.~R. 2012, \apj, 744, 140

\bibitem[{{Warren} {et~al.}(1995){Warren}, {Kallemeyn}, \&
  {Mayeda}}]{warren1995}
{Warren}, H.~P., {Kallemeyn}, W.~G., \& {Mayeda}, T. 1995, in Antarctic
  Meteorites XX, ed. K.~V. {Nayak}, Vol.~20, 261--264

\bibitem[{{Warren}(2005)}]{warren2005}
{Warren}, P.~H. 2005, Meteoritics and Planetary Science, 40, 477

\bibitem[{{Warren} \& {Kallemeyn}(1990)}]{warren1990}
{Warren}, P.~H., \& {Kallemeyn}, G.~W. 1990, LPSC 21, 1295

\bibitem[{{Wiechert} {et~al.}(2001){Wiechert}, {Halliday}, {Lee}, {Snyder},
  {Taylor}, \& {Rumble}}]{wiechert2001}
{Wiechert}, U., {Halliday}, A.~N., {Lee}, D.-C., {Snyder}, G.~A., {Taylor},
  L.~A., \& {Rumble}, D. 2001, Science, 294, 345

\bibitem[{{Wiik}(1972)}]{wiik1972}
{Wiik}, H.~B. 1972, Meteoritics, 7, 553

\bibitem[{{Zhang} {et~al.}(2012){Zhang}, {Dauphas}, {Davis}, {Leya}, \&
  {Fedkin}}]{zhang2012}
{Zhang}, J., {Dauphas}, N., {Davis}, A.~M., {Leya}, I., \& {Fedkin}, A. 2012,
  Nature Geoscience, 5, 251

\end{thebibliography}
\end{document}